\documentclass{article}
\usepackage{arxiv}

\usepackage{graphicx,url}
 \usepackage[utf8]{inputenc}
\usepackage[brazil]{babel}
 \usepackage{lettrine} 

\usepackage[dvipsnames]{xcolor}
\usepackage{todonotes} 

\usepackage{placeins}

\usepackage{caption}
\usepackage{subcaption}

\usepackage{graphicx}

\usepackage{pgfplots}
\usepgfplotslibrary{groupplots}
\usepackage{tikz}
\usetikzlibrary{patterns}
\pgfplotsset{compat=newest}
\usepackage{pgfplotstable}
\usetikzlibrary{patterns}
\pgfplotsset{every tick/.style={black,}}
\usepackage{amsmath}

\pgfplotsset{%
    compat=1.8,
    compat/show suggested version=false,
    tick label style={font=\footnotesize},
    label style={font=\footnotesize},
    legend style={%
        font=\footnotesize,
        at={(0.5,-0.15)},
        anchor=north,
        legend columns=2
        legend cell align=left,
        nodes={inner xsep=2pt,inner ysep=0.4pt,text depth=0.15em},
    },
}

\title{Experiências, Resultados e Reflexões a partir do Gerenciamento de experimentos no Mundo Real com FANETs e VANTs -- Versão Estendida}

\author{
 Bruno José Olivieri de Souza \\
  Department of Informatics\\
  Pontifícia Universidade Católica do Rio de Janeiro\\
  Rio de Janeiro, Brazil \\
  \texttt{bolivieri@inf.puc-rio.br} \\
  \And
 Markus Endler \\
  Department of Informatics\\
  Pontifícia Universidade Católica do Rio de Janeiro\\
  Rio de Janeiro, Brazil \\
  \texttt{endler@inf.puc-rio.br} \\
}

\begin{document} 

\maketitle
\begin{abstract} 
In the research on FANETs (Flying Ad-Hoc Networks) and distributed coordination of UAVs (Unmanned Aerial Vehicles), also known as drones, there are many studies that validate their proposals through simulations. Simulations are important, but beyond them, there is also a need for real-world tests to validate the proposals and enhance results. However, field experiments involving drones and FANETs are not trivial, and this work aims to share experiences and results obtained during the construction of a testbed actively used in comparing simulations and field tests.
\end{abstract}


\section{Introdução}
\label{sec:introducao}

Em um subconjunto significante de trabalhos acadêmicos de ciência da computação as propostas apresentadas quando não são provas formalmente, são amplamente verificadas através de testes e suas análises. Esses testes apresentam resultados a partir de \textit{datasets} previamente disponíveis em áreas de conhecimento como Banco de Dados e Aprendizado Supervisionado, ou, em grande parte, a partir de experimentos planejados e implementados em ambientes simulados pelos próprios proponentes do trabalho.

A realização de análises e \textbf{$V$}erificações de propostas através de simulação é algo extremamente difundido em diversas áreas de concentração como em redes, sistemas distribuídos, visão computacional, otimização dentre outras. O uso de simulações é amplamente acessível e geralmente pode ser utilizado por um único pesquisador. Isso habilita a pesquisa a evoluir de maneira mais ágil e não esbarra com os entraves de se adquirir equipamentos onerosos para avaliar uma única hipótese com o risco do equipamento não voltar se ser utilizado em outra pesquisa.

Por um lado as simulações agilizam e viabilizam cenários outrora não acessíveis, por outro lado, elas trazem uma simplificação de todas as variáveis de ambiente que poderiam agir sobre um experimento. Essa simplificação tende a crescer significativamente quando uma proposta é simulada com uma implementação do próprio pesquisador ao invés de utilizar \textit{frameworks} robustos, como por exemplo, NS3, OMNET++/INET no caso de propostas que tangenciem temas relacionados à redes de computadores. 

Além das \textbf{$V$}erificações também pode-se realizar \textbf{$V$}alidações em cenários reais. Dessa maneira inúmeros fatores ambientais, cenários e equipamentos passam a interferir nos experimentos de uma proposta em análise. Essas \textbf{$V$}alidações, ou experimentos realísticos\footnote{\url{https://youtu.be/rz02rnxYYQY?si=hISwZ3T3D89Gy1dZ}}, possibilitam um enorme, ou total, realismo e acoplamento com a realidade, fazendo com que os resultados sejam mais acurados. 

Este trabalho contribui com algumas experiências na construção, utilização e manutenção de um \textit{testbed} utilizado para ser utilizado em \textbf{$V$}alidações de propostas e trabalhos que incluem problemáticas de redes e sistemas distribuídos no contexto de FANETs (Flying Ad-Hoc Networks). Este \textit{tedbed}, chamado de GrADyS Framework é composto por simuladores, por VANTs autônomos (a.k.a. drones), composto também por redes de sensores efetivamente instaladas em campo, assim como outros equipamentos como \textit{rovers} autônomos (UGVs) que se locomovem no solo. Também são apresentados alguns resultados de aferição em campo em contraste com simulações e seus desdobramentos. O projeto GrADyS visa realizar \textbf{$V$}erificações simuladas e \textbf{$V$}alidações em campo para enriquecer as discussões de coordenação de enxames de drones\cite{bolivieri2023}.

O trabalho está segue organizado com a seção \ref{sec:contexto-ref-teorico} sobre detalhes do contexto de “\textbf{$V$}erificação e \textbf{$V$}alidação”; a seção \ref{sec:gradys-framework} apresenta o GrADyS framework; a seção \ref{sec:trab-relacionados} apresenta os trabalhos mais intrinsecamente relacionados; a seção \ref{sec:experiencias} apresenta lições aprendidas com experimentos de mundo real; a seção \ref{sec:resultados} apresenta resultados de experimentos em campo e a seção \ref{sec:discussoes-clonclusao} as conclusões.


\section{Verification $\&$ Validation}
\label{sec:contexto-ref-teorico}

O termo “Verificação e Validação” (a.k.a. \textbf{$V$}$\&$\textbf{$V$}) é utilizado para o acoplamento de dois processos importantes em pesquisas e refere-se a práticas importantes na busca de qualidade e confiabilidade dos resultados de uma pesquisa. 

\subsection{\textbf{$V$}erificação} 
A \textbf{$V$}erificação é um processo que visa garantir que uma proposta científica é válida conforme os aspectos de suas suposições e limitações intrínsecas. Usualmente o pesquisador recorre a uma implementação própria ou se vale de simuladores os quais implementam um conjunto de variáveis de ambiente ou de cenários. Envolve verificar e confirmar se a proposição e a implementação da proposta estão conformes com as suposições ou produzem dados para que possam ser analisadas e discutidas. As atividades de \textbf{$V$}erificação podem incluir muitas simulações, inspeções, revisões, orientações e outros métodos estáticos para detectar erros no início do processo de análise. 

Algumas pesquisas seriam muito perigosas ou mesmo impraticáveis de serem realizadas, principalmente no âmbito de pesquisa básica. Porém, as \textbf{$V$}erificações também são amplamente utilizadas em pesquisas no âmbito da pesquisa aplicada procurando redução de custos e agilidade. Isso é muito notório na área de redes de computadores, pois, mesmo havendo diversos equipamentos acessíveis como smartphones dotados de 2, 3, em alguns casos quatro ou cinco rádios embarcados, algumas propostas científicas são avaliadas apenas com simulações.

\subsection{\textbf{$V$}alidação}

O processo de \textbf{$V$}alidação em pesquisa acadêmica é crucial para garantir a confiabilidade e a credibilidade dos resultados obtidos. A \textbf{$V$}alidação visa confirmar se os métodos utilizados são apropriados, se os resultados são confiáveis e se as conclusões são válidas. 

Usualmente a \textbf{$V$}alidação transpassa a \textbf{$V$}eriricação através de simulações ou ambientes estanques para que a proposta acadêmica seja analisada em um ambiente operacional realístico. Isso faz com que os resultados coletados para análise sejam mais sólidos ao mesmo tempo que trazem a dificuldade de lidar com o ambiente real seja por logística ou o risco do cenário ser muito particular, algo que pode ser facilmente evitado em simulações.

\subsection{\textbf{$V$}erificação e \textbf{$V$}alidação}

Em resumo, a \textbf{$V$}erificação centra-se nas fases de design e implementação para a análise da proposta da pesquisa ainda em construção, enquanto a \textbf{$V$}alidação centra-se nas fases de teste e avaliação para garantir que o produto de uma pesquisa possa ser construído e testado frente as propostas e abordagens relacionadas.

Ambos os processos são essenciais para a produção de resultados confiáveis e de alta qualidade em pesquisa acadêmica. Eles são partes integrantes do rigor científico, contribuindo para a validade interna e externa dos estudos.

\section{GrADyS Framework}
\label{sec:gradys-framework}

\begin{figure}[h!]
  \caption{A metodologia do Projeto GrADyS ilustrando seus 4 componentes de simulação no laboratório e em campo (\textit{a.k.a in the wild}).}
  \label{fig-gradysmetodology}
  \centering
  \includegraphics[width=0.95\textwidth]{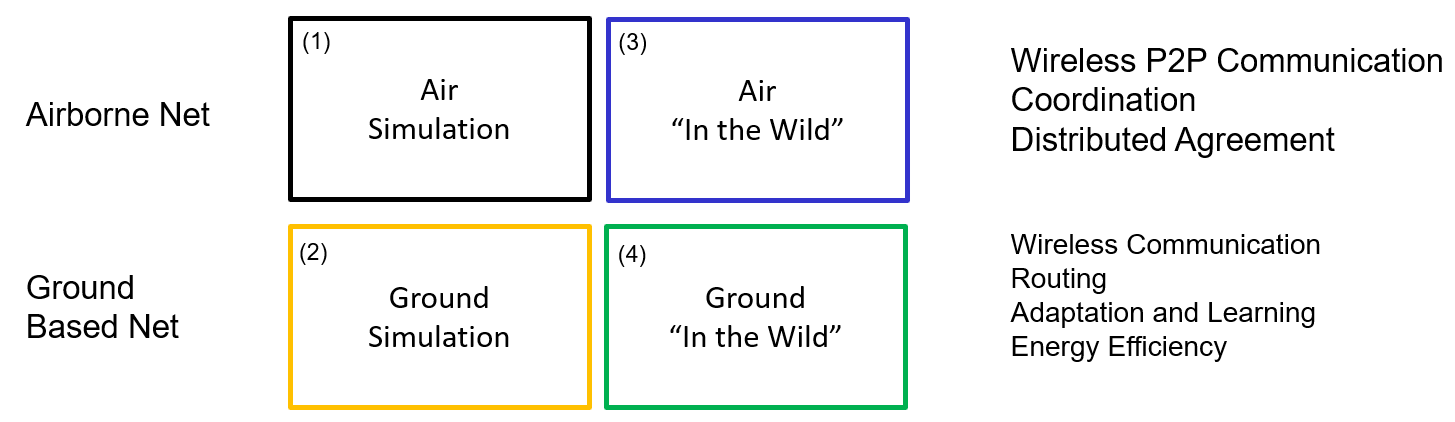}
\end{figure}

O GrADyS Framework consiste na união de diversos componentes, sendo sua maior força a capacidade de \textbf{$V$}alidar no mundo real as simulações realizadas com drones e, em alguns casos, redes de sensores sem fio (RSSF). A Figura \ref{fig-gradysmetodology} ilustra a abordagem de trabalho do projeto.

O framework pode ser observado em seus quatro pilares:(1) Simulação de protocolos e algoritmos distribuídos para coordenação de nós móveis com comunicação ad hoc; (2) Simulações de RSSF e algoritmos de roteamento; (3) A coordenação de VANTs em voos reais e (4) A efetiva implementação de uma RSSF com sensores no solo.

A simulação de protocolos e algoritmos distribuídos para coordenação de nós móveis é habilitada pela utilização de um mesmo código de coordenação feito em Python para ser executado de maneira simplificada para a rápida prototipação em simulações e em veículos. Ao mesmo tempo o mesmo trabalho pode fazer uso da suíte OMNET++/INET para que execute suas comunicações de maneira mais fidedigna e coordena a dinâmica de movimentação de veículos aéreos em um simulador SITL (\textit{software in the loop}) de controladores de voos reais baseados em Ardupilot\footnote{\url{https://ardupilot.org/}}. Resultados específicos nesta linha podem ser analisados no trabalho GrADyS-SIM\cite{gradysSIMsbrc2022} baseado em OMNET++/INET, no MAVSIMNET\footnote{\url{https://thlamz.github.io/MAVSIMNET/}} que promove o bind entre o protocolo Mavlink e o OMNET++/INET, assim como diversos outros pontos disponíveis na página do projeto\footnote{\url{https://www.lac.inf.puc-rio.br/index.php/gradys/}} e em seu blog informal\footnote{\url{https://gradys.tumblr.com/}}.

As simulações de RSSF e algoritmos de roteamento são implementadas herdando as pilhas de implementação presentes no OMNET++/INET nas quais implementações BLE e 802.15.4 tem seus resultados apresentados trabalhos prévios\cite{paulon2022}.

A efetiva implementação de RSSF no solo talvez seja umas das mais escassas \textbf{$V$}alidações em trabalhos relacionados. Em nossos trabalhos alteramos a pilha de implementação nativa de controladores baseados no chip ESP32 para prover algoritmos de roteamento BLE com uso mais eficiente de energia do que a especificação do próprio BLE\cite{paulonEsp32}. 

A coordenação de VANTs em voos reais é realizada com quadricópteros de baixo custo e baixa sofisticação, cujas peças são adquiridas individualmente e são montados pelos alunos. Os veículos são equipados com \textit{single board computers} (SBC), usualmente RaspBerry Pis, que habilitam a codificação dos protocolos e algoritmos distribuídos para coordenação ao mesmo tempo que possuem rádios 802.11 e 802.15.4.  \textbf{$V$}erificação com veículos reais habilita todo um novo nível de confiança nos testes ao mesmo tempo que expões novos desafios para rodar os testes.

\begin{figure}[h!]
  \caption{O framework GrADyS ilustrando seus componentes de simulação no laboratório e seus componentes em campo.}
  \label{fig-gradysframework}
  \centering
  \includegraphics[width=0.95\textwidth]{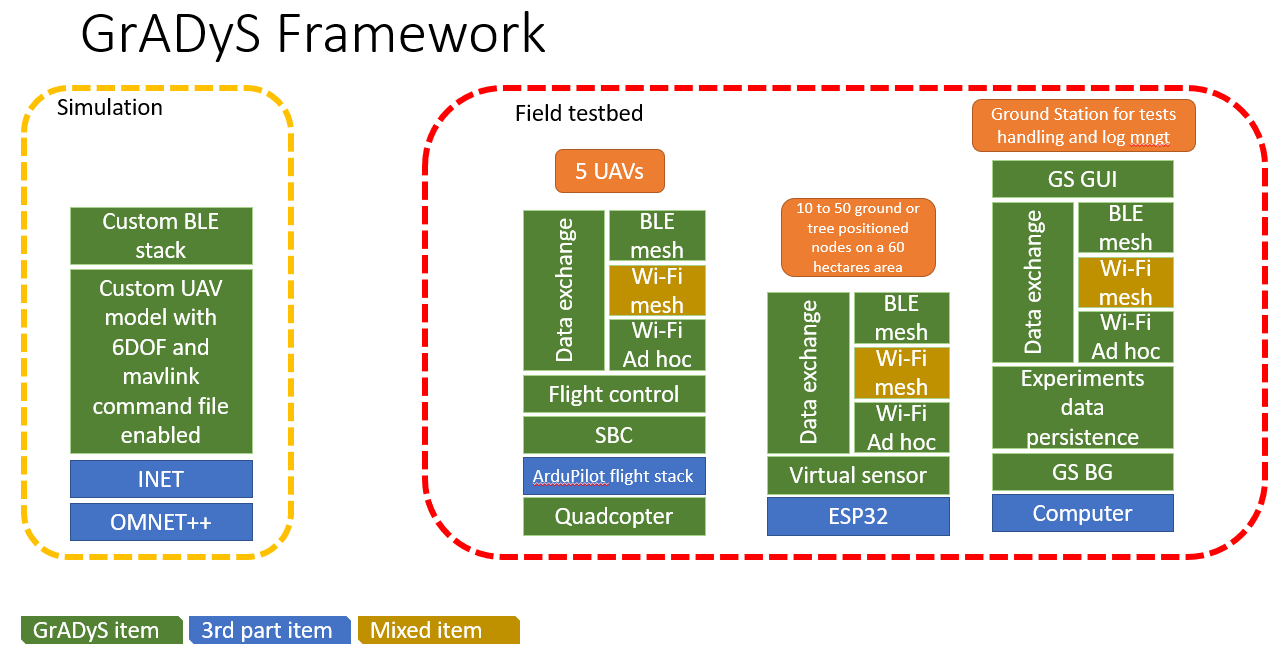}
\end{figure}

No mundo real a implementação do framework é realizada por diversos componentes ilustrados na Figura \ref{fig-gradysframework}. Nela são apresentados os componentes de simulações e os componentes de testes em campo com as seguintes ferramentas:

\begin{itemize}
    \item Ferramentas de simulação:
        \subitem GrADyS-SIM\cite{gradysSIMsbrc2022}
        \subitem GraDys-SIM NG\cite{gradys-sim-ng-2024}
    \item Ferramentas utilizadas no campo:        
        \subitem GrADyS GS - Ground Station\cite{Perricone2022}
        \subitem Altered Bluetooth routing strategies\cite{paulon2022}\cite{paulon2023}
        \subitem QuadcopterPI software\footnote{https://github.com/brunoolivieri/quadpi}
        \subitem Mavlink protocol stack\footnote{https://mavlink.io/en/}
        \subitem ArduPilot software stack\footnote{https://ardupilot.org/}
\end{itemize}

\section{Trabalhos relacionados}
\label{sec:trab-relacionados}

Na pesquisa de protocolos de coordenação de enxames de VANTs, diversos trabalhos implementam simuladores para apresentar e verificar suas pesquisas. Podem se destacar trabalhos como o AirSim\cite{airsim2018}, FlyNetSim\cite{FlyNetSim2018}, GrADyS-SIM\cite{gradysSIMsbrc2022}, ArduSim\cite{ardusim2018} e diversos outros.

A larga maioria de trabalhos focam em simulação e, usualmente, a parcela ligada a comunicação entre nós é menos fidedigna e simplificada. Exceções à esta afirmação, no melhor de nosso conhecimento, é presente quando há alguma sinergia com ambientes ROS e no MAVSIMNET\footnote{\url{https://thlamz.github.io/MAVSIMNET/}}. 

A implementação de \textit{testbeds} reutilizáveis em mundo ral para a \textbf{$V$}alidação de propostas com VANTs tem pouco mais de uma década com o trabalho de Vijay Kumar et al \cite{Kumar2010} na UnPen. Nesse trabalho são utilizados vários micro VANTs que se comunicam com um nó central controla a todos com um modelo centralizado em MatLab. Todo o sistema de posicionamento é \textit{indoor} a viabilizado com sistemas de capturas de imagem\cite{simulEnvs2018}, como, por exemplo as VICON\footnote{\url{https://www.vicon.com/}}. Na sequência desse trabalho, outros \textit{testbeds} reutilizáveis com a mesma arquitetura foram sendo implementados pela academia como no ETH por Rafaello D’Andrea et al 2014  \cite{lupashin2014platform} ou mesmo com veículos sub aquático como por Sidney et al. \cite{Sydney:2010:MTU:2377576.2377597}.

Mais de uma década depois, novos trabalham são apresentados com arquitetura semelhante como por Guerreiro et al. \cite{Guerreiro2021} no qual a diferença marcante é que o sistema de posicionamento dos VANTs é controlado por UWB (\textit{ultra wide band}) e o controle permanece centralizado, emulando a independência entre os nós e problemas em FANETs.

Nos casos citados acima há sempre a carência de testes outdoor que invariavelmente são mais difíceis de gerenciar quanto a interferências, sejam ambientais ou de telecomunicações ou apenas de logística. Outro fator, este sim, perpendicular a todos os trabalhos mencionados é quanto ao processamento dos nós emulados ou veículos reais que frequentemente é realizado em um nó central com alto poder de processamento. Em nosso trabalho, os algoritmos distribuídos são efetivamente testados em nós separados que se comunicam em uma FANET totalmente Ad Hoc.

\section{Experiências no mundo real}
\label{sec:experiencias}

Mesmo em grupos de pesquisa mais maduros ou em laboratórios temáticos com diversos pesquisadores em departamentos universitários robustos, muitas vezes grande parte da pesquisa ocorre compartimentada ou até mesmo individualizada. Em torno de um mesmo tema, diversos pós-graduandos e jovens pesquisadores realizam suas pesquisas e implementam suas simulações individualmente. Isso se torna prático em termos de \textbf{$V$}erificações. Porém, isso \textit{pode} dificultar o reaproveitamento e a reprodutibilidade das pesquisas.

\begin{figure}[h!]
\caption{Quatro drones decolando sincronamente.}
  \label{fig-drones}
  \centering
  \includegraphics[width=0.95\textwidth]{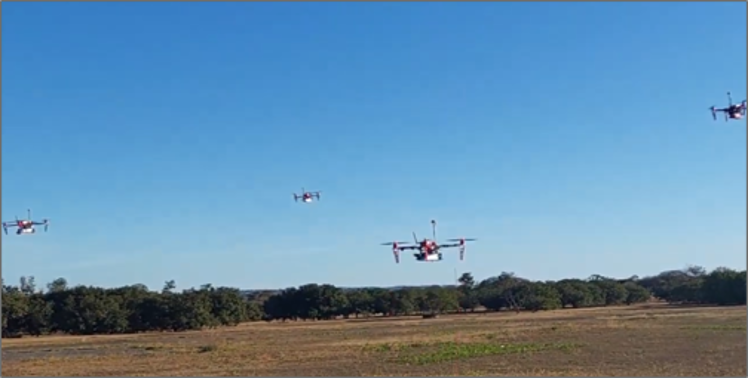}
\end{figure}

Quando começamos a falar em \textbf{$V$}alidações, esse cenário encontra dois principais obstáculos: o primeiro é que os custos atrelados ao migrar de simulações para teste no mundo real sempre, ou quase sempre, direcionam times de pesquisa a se indagar como reaproveitar os investimentos realizados e; em segundo lugar, um pesquisador sozinho dificilmente conseguirá implementar executar experimentos em mundo real com diversos equipamentos como na Figura \ref{fig-drones}\footnote{Vídeo em \url{https://youtu.be/VCO6h4jAAQk?si=SzdE1psVXQYvac-m}}.

\subsection{Sair do laboratório}
\label{sec-sub:sair-do-lab}

O primeiro passo geralmente é o mais difícil. Adquirir equipamentos consome tempo para um planejamento coerente e mais tempo ainda para a obtenção de financiamento adequado que, em muitas vezes, é insuficiente.

No caso do \textit{testbed} GrADyS descrito na seção \ref{sec:gradys-framework} há drones dotados de pelo menos quatro rádios o que trás a necessidade de logística significante. Além dos drones em si, são necessários diversos equipamentos como notebooks de apoio, antenas, infraestrutura de apoio e espaço, sendo esses dois últimos, merecedores de atenção especial. 

O espaço em alguns \textit{campi} universitários podem ser inexistentes e em outros, podem ser afastados e ermos. Para ensaios com drones e RSSF como descritos anteriormente, nosso \textit{testbed} possui uma área útil de 60ha. Em nossos experimentos possuíamos sessenta nós baseados em ESP32 para espalhar pelo campo te de testes, porém a (re)disposição ou troca de baterias nos locais exatamente iguais aos simulados é incrivelmente árdua. Na prática, nossos testes mais eficazes e eficientes foram realizados sobre uma área de 10ha. Este cenário demorada aproximadamente uma hora para ser (re)habilitado. 

Sem uma infraestrutura básica como mesas e tomadas, tudo fica mais difícil. Enquanto realizar testes \textit{outdoor} em dias de chuva torna-se inviável, dias ensolarados são igualmente desafiadores. O sol nas telas de notebooks as torna inutilizáveis e acessórios como guarda chuvas, caixas ou gazebos são primordiais. Mesas móveis, grandes extensões e \textit{walktalkies} são imprescindíveis em testes \textit{outdoor}. Diversas imagens desses momentos podem ser acessadas em \url{https://gradys.tumblr.com/}

\subsection{Veículos para uma FANET}
\label{sec-sub:exp-veiculos}

Veículos por sua natureza são inseridos em experimentos para se movimentar e drones levam isso à para um ambiente 3D, ao ar livre eles ficam bem suscetíveis à ventos e obstáculos perigosos. Preparar os veículos requer tempo e, ao menos, uma pessoa por veículo, se forem necessário dez minutos para habilitar um drone e forem  seis drones, ou se tem seis pessoas ou se gasta mais de uma hora.

\begin{figure}[h!]
\caption{Dois drones montados por alunos com RaspBerry Pis}
  \label{fig-dois-drones}
  \centering
  \includegraphics[width=0.50\textwidth]{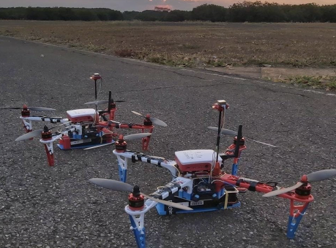}
\end{figure}

A decisão sobre a aquisição de drones prontos ou a montagem de protótipos próprios (como na Figura \ref{fig-dois-drones} pode ser difícil. Tomando como exemplo testes em mundo real de uma FANET\cite{fanetTestBed2015}, os drones são apenas os nós móveis que não fazem parte da contribuição principal da pesquisa, contudo, são extremamente necessários. Mais ainda, um grupo de pesquisa que visa a contribuição científica em FANETs talvez não possua o \textit{know-how} de montagem, manutenção e pilotagem ou mesmo automação de drones.

Esses fatores usualmente contribuem para aquisição de equipamentos de consumo, prontos para uso e neste mercado a marca DJI\footnote{\url{www.dji.com}} domina. Esses equipamentos possuem uma baixíssima barreira de entrada e, em um primeiro momento, se mostram a escolha mais ágil. Contudo eles não são concebidos para não carregarem nenhum \textit{payload} ou viabilizam um SDK habilitado que permita alguma customização ou automação significante. 

Antes da decisão de que drones ou demais veículos serão utilizados, uma análise mais profunda sempre é importante para que possam entender as expectativas não iniciais e o investimento possa ser reaproveitado ou ser duradouro. A marca DJI continua sendo uma excelente opção desde que em família de drones bem mais robustas que tenham integração dom ROS\footnote{The Robot Operating System - \url{https://www.ros.org/}}. Veículos DIY com com projetos de OpenHardware e OpenSoftware fatalmente terão um tempo maior de maturação, porém serão largamente reaproveitados e customizáveis aos experimentos. Nesta ultima linha, o conjunto de software ArduPilot\footnote{\url{https://ardupilot.org/}} é robusto a atende virtualmente quaisquer tipo de veículos e possui grande padronização de APIs.

Para utilização de drones a logística antes, durante e depois dos voos não pode ser subestimada, muito tempo e esforço são gastos. São necessárias várias atividade em paralelo, desde a pilotagem ou monitoramento de voo autônomo em segurança, o controle da área de testes para garantir segurança e a coleta de dados observada. Um simples erro na coleta dos dados detectado depois da atividade em campo pode fazer com que todo o esforço tenha sido perdido e tudo teria que ser feito novamente.

\subsection{Baterias}
\label{sec-sub:exp-baterias}

Drones utilizam baterias que invariavelmente são confeccionadas com Lítio\footnote{\url{https://en.wikipedia.org/wiki/Lithium_battery}}. Em sua maioria, são baterias de LiPo ou LiOn tal como as baterias de carros elétricos. Essas baterias tem como vantagem, o alto poder de descarga. De maneira direta essas características fazem com essas baterias sejam de difícil aquisição, pois há diversas restrições de transporte e estocagem. Esses pontos se traduzem em alto custo.

Uma bateria pode ser carregada em até uma hora à uma taxa de carga de 1C e deve prover um voo de dez ou, no máximo, 20 minutos. Isso traz mais dois desafios: (1) são necessárias muitas baterias para um bom dia de testes e (2) para que as baterias possam ser carregadas em tempo razoável, é necessário paralelizar o carregamento. Um ponto desafiador é que baterias de Lítio não devem ser armazenadas totalmente carregadas nem descarregadas.

Essas baterias são perigosas, pois podem pegar fogo e até explodir. O manuseio incorreto de baterias LiPo pode causar incêndio, explosões e inalação de fumaça tóxica. Deve-se planejar o uso das baterias com cautela e seguir recomendações de armazenamento rigorosamente. As atividades para carregar as baterias devem sempre ser observadas.

\subsection{Rádios}
\label{sec-sub:exp-radios}

Uma vez vencidos os obstáculos referentes aos veículos da FANET, os rádios passam a ser um ponto importante. Dentre várias opções de chips e dispositivos, é possível criar dois grandes grupos. Primeiro, com forte ligação com os microcomputadores, há uma infinidade de rádios que habilitam 802.11 e por sua vez redes IP. Em segundo há rádios com \textit{stacks} mais leves e semelhantes entre sí como 802.15.4, LoRa, ESP-NOW, ZigBee dentre outras\cite{loraVS8021542020}.  

Para aplicações que fazem troca de dados estruturados ou volumosos é interessante utilizar rádios que deem suporte à alguma versão do 802.11. Para pequenas trocas de inteiros ou poucos dados, deve se priorizar os rádios 802.15.14. Embora essas afirmações pareçam diretas a partir da natureza dois padrões citados elas vem da experiência na utilização dos rádios que as suportam. As \textit{stacks} e APIs\footnote{\url{https://en.wikipedia.org/wiki/API}} de ambos os casos são diametralmente opostas nas ofertas de funcionalidades e forma de utilização.

Se por um lado, usar o padrão 802.11 lhe oferece quase incontáveis equipamentos e bibliotecas de implementação interoperáveis, por outro lado, o padrão 802.15.4 tem oferta escassa de bibliotecas e seus subpadrões (LoRa, ESP-NOW, ZigBee, etc) fragmentam a oferta de hardware e restringem muito a interoperabilidade. Rádios que se baseiam  no padrão 802.15.4 irão permitir uma comunicação muito mais simples e com maior alcance do que rádios entregam o padrão 802.11 e isso deve ser ponderado.

\subsection{Antenas}
\label{sec-sub:exp-antenas}

Após falar dos rádios, a utilização das antenas merece uma atenção redobrada. Muitos equipamentos trazem antenas impressas diretamente em suas placas, tal como alguns ESP32 e o RaspBerry Pi. Essas antenas usualmente funcionam muito bem em pequenas distâncias, mas quando esses equipamentos são instalados em drones, o alcance fica restrito a poucos metros, inviabilizando quaisquer testes. Vale ratificar que os motores dos drones geram muito ruído no circuito e filtro de banda baixa devem ser utilizados.

Sempre devem ser priorizadas antenas externas aos equipamentos e de boa qualidade. O mercado está inundado de antenas com o corte errôneo para o comprimento de onda almejado. De fato, em nosso testbed os drones usando RaspBerry Pi com antena original era acessível no máximo a 10 metros quando em voo. Utilizando uma antena e rádio USB conseguimos alcançar 40 metros. Por fim, o rádio original do Raspberry Pi tendo sua antena padrão removida e uma nova antena modificada instalada chegou aos 200 metros de alcance, o máximo esperado para o padrão 802.11.

Outro ponto importante sobre antenas que não pode ser subestimado é seu posicionamento. É esperado, e nós testamos, que dois simples ESP32 usando 802.15.4 consigam trocar dados a 180 metros quando colocados suas antenas estejam paralelamente entre si, perpendiculares ao solo e distante do solo a 2 metros. Caso as antenas sejam colocadas deitadas perto do solo a uns 10cm, esse alcance de 180 metros é reduzido à 40cm. Isso é totalmente esperado dada a irradiação do sinal.

\subsection{Companion Computers}
\label{sec-sub:exp-soc-sbc}

Para experiências quem envolvam sistemas distribuídos, os nós devem ser logicamente independentes e, no caso de enxames de drones executando algoritmos efetivamente distribuídos, são necessários nós de processamento embarcados. Esses nós deverão possuir os rádios e antenas supracitados.

Considerando que o drone será responsável somente pelo processamento de controle do voo, o experimento deve ser executando no \textit{companion computer} que será embarcado no drone. Para fazer esse papel de \textit{companion computer} é usual que sejam ponderadas duas opções: (1) System on a Chip (SoC) tal como ESP32 e outros $\&$ (2) Single Board Computers (SBC) tal como Raspberry Pi, NVidia Xavier e outros. Ambos tem vantagens e desvantagens:

\begin{enumerate}
    \item[i] SoC: 
        \begin{itemize}      
            \item Vantagens: São bem pequenos e fáceis de embarcar, são leves e baratos. Também são mais resistentes do que que SBC e trabalham bastante tempo com baterias;
            \item Desvantagens: Pouco poder de processamento, de fato, deve se atentar que os núcleos de processamento lógico do código inserido pode ser o mesmo que irá processar os pacotes de rádio e isso pode interferir. Alguns paradigmas de programação não estão disponíveis, como por exemplo, multiprocessamento. O código que será inserido usualmente será inserido de maneira manual (instalar o \textit{firmware}), isso se tona muito pouco prático e de difícil customização. Há pouco reaproveitamento de outras pesquisas e projetos com código aberto;
        \end{itemize}
        
    \item[ii] SBC:
        \begin{itemize}      
            \item Vantagens: Bom poder de processamento, bastante memória, múltiplos rádios embarcados (ao menos WiFi e BLE, altamente expansível, uso de Sistema Operacionais e demais possibilidades tal como um computador completo, tal como um laptop. Há muito reaproveitamento de outras pesquisas e projetos com código aberto;
            \item Desvantagens: São mais caros. De fato, um RaspBerry 4 com 4gb na data atual custa o mesmo valor de um mini PC com processador intel N5095 que possui muito mais recursos, mas que não é facilmente embarcável. SBCs não trabalham tanto tempo com baterias como os SoC e nem sempre são pequenos e leves para serem embarcados.
        \end{itemize}

    \item[iii] Tradeoffs: Embora os SoC habilitem um começo rápido e \textit{"quick wins"}, eles são muito escassos em recursos. Isso fica mais difícil de gerenciar fora do cerne do experimento, nas soluções que rodam em paralelo coletando dados, armazenando logs e os organizando. Em nossas experiências, mesmo que estivéssemos focando em um SoC Nordic\footnote{\url{https://www.nordicsemi.com/products/nrf52840}} ou ESP32, fatalmente era necessário com SBC entre ele o drones para gerenciar o experimento, coletar dados e controlas o drone. Além disso, o uso de um SBC embarcado habilita instantaneamente um conjunto muito grande de ferramentas em cima do sistema operacional tal como o uso de Python, ROS, ROS2, bancos de dados e brokers de mensageria.

\end{enumerate}





\subsection{Logs e sincronização}
\label{sec-sub:exp-logs-sync}

Uma questão que gasta muito tempo em pesquisas de redes e em sistemas distribuídos é o \textit{debuging} dos código que muita vezes roda em algum tipo de assincronicidade ou paralelismo. Para trabalhar este ponto muitas IDEs trazem recursos para observar valores em paralelo e programadores geram uma quantidade significativa de logs. 

Contudo, em experimentos reais há um forte fator dificultador que não ocorre nas simulações. Os SoC e SBC invariavelmente não possuem um Real Time Clock (RTC\footnote{\url{https://en.wikipedia.org/wiki/Real-time_clock}}) com bateria a bordo. Com isso, mesmo seu sejam utilizados cartões de memória para mecanismos de log, ainda haverá a dificuldade de sincronizá-los. SBC costumam sincronizar seus relógios via Internet e, em uma FANET, isso pode não estar disponível. RTCs podem ser manualmente instalados, mas geralmente não são muito precisos e requer mais customização na eletrônica do drone.

Uma solução que adotamos foi acessar as APIs dos drones para resgatar o horário presente nos GPS, esses sim, extremamente precisos. Esta solução embora pareça trivial, é trabalhosa. 

Em todos os casos, mesmo com logs sincronizados nos nós, eles estão espalhados em cada nó, em cada drone ou sensor. Resgatá-los pode ser demorado como o processo de carga das baterias. Para isso uma Ground Stations (GS) é algo útil e pode coletar os dados ainda em campo para uma verificação se estão íntegros e o experimento realmente vai poder ser analisado fora do campo. A gestão de vários SD cards e arquivos é temorosa. 

\subsection{Ground Stations}
\label{sec-sub:exp-groundstations}

Por fim, sempre que há um experimento, há uma coleta de dados. Em nossas experiências passadas, usualmente cada aluno começava de maneira independente construindo um par de scripts que culminavam em um belo conjunto de código e análises, isso quando não viravam Ground Stations (GS) completas para controlar os drones e os experimentos.

Esse trabalho pode ser melhor aproveitado se planejado e reaproveitado em mais de uma linha de pesquisa. Nesse pensamento, o mais usual é querer observar o que está acontecendo, coletar dados e enviar comandos para um subconjunto ou todos os drones. 

Pensando nisso, nosso projeto implementou e disponibilizou o GrADyS-GS\cite{Perricone2022}\footnote{\url{https://github.com/Project-GrADyS/gradys-gs}}. Trata-se de uma GS \textit{web-based} de fácil manutenção e customização implementada em Python e JS focando na simples troca e armazenamento de JSONs entre os veículos e a GS. Com ela múltiplas pessoas podem observar os mesmo experimento em terminais diversos e ao mesmo tempo controlar partes dele de maneira colaborativa.




\section{Resultados e lições aprendidas}
\label{sec:resultados}

Seguindo a linha de \textbf{$V$}alidar nossas propostas e simulações, abaixo apresentamos três experiências e algumas lições aprendidas. Em todos os nossos experimentos são explorados paradigmas sem infraestrutura centralizada para as comunicações, sendo a GS apenas um nó Ad Hoc, porém estacionário.

\subsection{Redes Mesh - 802.11s}
\label{sec-sub:resultados-mesh}

\begin{figure}
     \centering
     \begin{subfigure}[t]{0.45\textwidth}
         \centering
         \includegraphics[width=\textwidth]{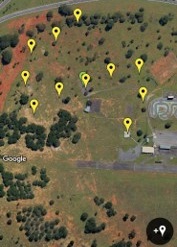}
         \caption{Mapa dos sensores dispostos em 10ha}
         \label{fig-sensores-e-mapa-mapa}
     \end{subfigure}
     \hfill
     \begin{subfigure}[t]{0.45\textwidth}
         \centering
         \includegraphics[width=\textwidth]{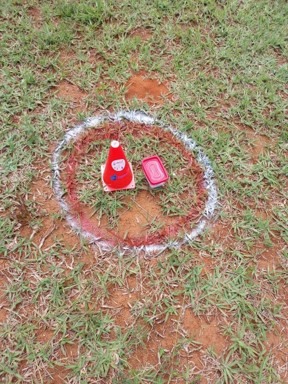}
         \caption{Exemplo de sensor no solo}
         \label{fig-sensores-e-mapa-sensor}
     \end{subfigure}
    \caption{Three simple graphs}
  \label{fig-sensores-e-mapa}
\end{figure}

O teste consistia em voar com drones sobre um campo e coletar dados de 10 sensores no solo e quantificá-los. Nesses experimentos em cada drone era utilizado um RaspBerry Pi apenas para controlar os drones e um ESP32 como rádio e para rodar o teste em si. No solo cada sensor era montado com um ESP32 com podem ser vistos na Figura \ref{fig-sensores-e-mapa}. Um teste muito simples para se testar os componentes do ambiente.


A comunicação entre os nós foi utilizando os rádios dos ESP32 como uma rede mesh, utilizando o PainlessMesh\footnote{\url{https://gitlab.com/painlessMesh/painlessMesh}} com o rádio configurado para 11 dBm. Dessa maneira esperava-se trocar dados via JSON entros drones e também com o nós no solo. 

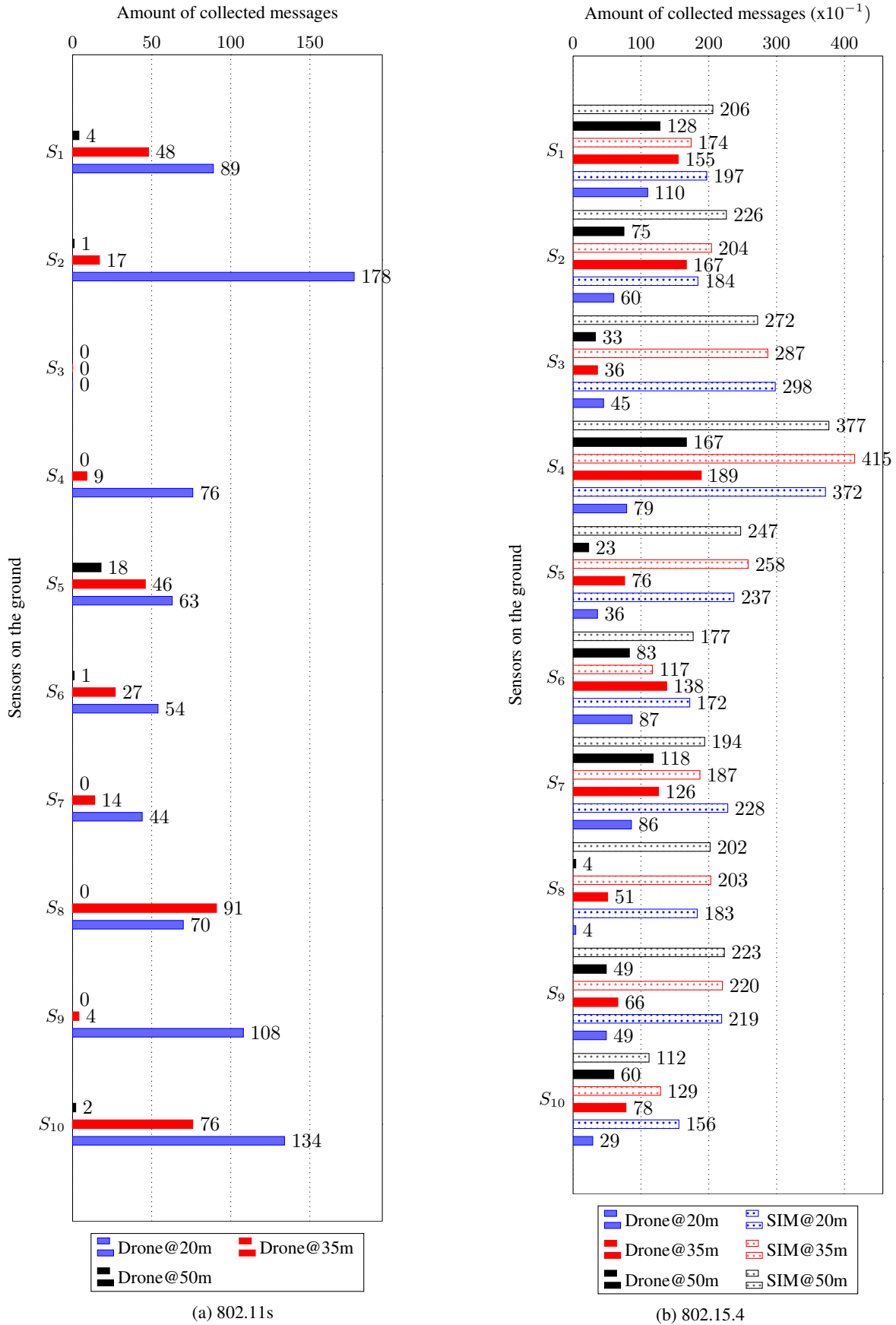
\begin{figure}
\centering
\begin{subfigure}[t!]{0.49\textwidth}
\begin{tikzpicture} 
    \begin{axis}[
        height=22cm, width=7cm,
        xbar, xmin=0,
        xlabel={Amount of collected messages},
        ylabel={Sensors on the ground},
        symbolic y coords={$S_{10}$,$S_9$,$S_8$,$S_7$,$S_6$,$S_5$,$S_4$,$S_3$,$S_2$,$S_1$},
        ytick=data,
        nodes near coords,
        nodes near coords align={right},
        point meta=rawx,
        xbar=4pt,
        bar width=0.15cm,
        tickwidth=0pt,
        xmajorgrids=true,
        major grid style={dotted,black},
        axis x line*=top,
        legend style={
        	at={(0.5,-0.01)},
        	anchor=north,
        	legend columns=2,
        	row sep = 5pt,
        	/tikz/every even column/.append style={column sep=0.4cm}
        },
    ]
    \draw (axis description cs:0,0) -- (axis description cs:1,0); 
    \addplot [draw = blue, pattern color = blue, fill = blue!60] coordinates {
      (89,$S_1$)
      (178,$S_2$)
      (0,$S_3$)
      (76,$S_4$)
      (63,$S_5$)
      (54,$S_6$)
      (44,$S_7$)
      (70,$S_8$)
      (108,$S_9$)
      (134,$S_{10}$) };
    \addlegendentry{Drone@20m}
    \addplot [draw = red, pattern color = red, fill = red] coordinates {
      (48,$S_1$)
      (17,$S_2$)
      (0,$S_3$)
      (9,$S_4$)
      (46,$S_5$)
      (27,$S_6$)
      (14,$S_7$)
      (91,$S_8$)
      (4,$S_9$)
      (76,$S_{10}$) };
    \addlegendentry{Drone@35m}
    \addplot [draw = black, pattern color = black, fill = black] coordinates {
      (4,$S_1$)
      (1,$S_2$)
      (0,$S_3$)
      (0,$S_4$)
      (18,$S_5$)
      (1,$S_6$)
      (0,$S_7$)
      (0,$S_8$)
      (0,$S_9$)
      (2,$S_{10}$) };
    \addlegendentry{Drone@50m}
  \end{axis}
\end{tikzpicture}
\caption{802.11s}
\label{fig:result_5ms_painlessmesh}
\end{subfigure}
\begin{subfigure}[t!]{0.49\textwidth}
\centering
\begin{tikzpicture} 
    \begin{axis}[
        height=21.5cm, width=7cm,
        xbar, xmin=0,
        xlabel={Amount of collected messages (x$10^{-1})$},
        ylabel={Sensors on the ground},
        symbolic y coords={$S_{10}$,$S_9$,$S_8$,$S_7$,$S_6$,$S_5$,$S_4$,$S_3$,$S_2$,$S_1$},
        ytick=data,
        nodes near coords,
        nodes near coords align={right},
        point meta=rawx,
        xbar=4pt,
        bar width=0.15cm,
        tickwidth=0pt,
        xmajorgrids=true,
        major grid style={dotted,black},
        axis x line*=top,
        legend style={
        	at={(0.5,-0.01)},
        	anchor=north,
        	legend columns=2,
        	row sep = 5pt,
        	/tikz/every even column/.append style={column sep=0.4cm}
        },
    ]
    \draw (axis description cs:0,0) -- (axis description cs:1,0); 
    \addplot [draw = blue, pattern color = blue, fill = blue!60] coordinates {
      (110,$S_1$) 
      (60,$S_2$)
      (45,$S_3$)
      (79,$S_4$)
      (36,$S_5$)
      (87,$S_6$)
      (86,$S_7$)
      (4,$S_8$)
      (49,$S_9$)
      (29,$S_{10}$) };
    \addlegendentry{Drone@20m}
    \addplot [draw = blue, pattern = dots, pattern color = blue] coordinates {
      (197,$S_1$)
      (184,$S_2$)
      (298,$S_3$)
      (372,$S_4$)
      (237,$S_5$)
      (172,$S_6$)
      (228,$S_7$)
      (183,$S_8$)
      (219,$S_9$)
      (156,$S_{10}$) };
    \addlegendentry{SIM@20m}
    \addplot [draw = red, pattern color = red, fill = red] coordinates {
      (155,$S_1$)
      (167,$S_2$)
      (36,$S_3$)
      (189,$S_4$)
      (76,$S_5$)
      (138,$S_6$)
      (126,$S_7$)
      (51,$S_8$)
      (66,$S_9$)
      (78,$S_{10}$) };
    \addlegendentry{Drone@35m}
    \addplot [draw = red, pattern = dots, pattern color = red!60] coordinates {
      (174,$S_1$)
      (204,$S_2$)
      (287,$S_3$)
      (415,$S_4$)
      (258,$S_5$)
      (117,$S_6$)
      (187,$S_7$)
      (203,$S_8$)
      (220,$S_9$)
      (129,$S_{10}$) };
    \addlegendentry{SIM@35m}
    \addplot [draw = black, pattern color = black, fill = black] coordinates {
      (128,$S_1$)
      (75,$S_2$)
      (33,$S_3$)
      (167,$S_4$)
      (23,$S_5$)
      (83,$S_6$)
      (118,$S_7$)
      (4,$S_8$)
      (49,$S_9$)
      (60,$S_{10}$) };
    \addlegendentry{Drone@50m}
    \addplot [draw = black, pattern = dots, pattern color = black!60] coordinates {
      (206,$S_1$)
      (226,$S_2$)
      (272,$S_3$)
      (377,$S_4$)
      (247,$S_5$)
      (177,$S_6$)
      (194,$S_7$)
      (202,$S_8$)
      (223,$S_9$)
      (112,$S_{10}$) };
    \addlegendentry{SIM@50m}
  \end{axis}
\end{tikzpicture}
\caption{802.15.4}
\label{fig:result_5ms_802154}
\end{subfigure}
\caption{Testes de FANETs com 802.11s e 802.15.4}
\end{figure}

A Figura \ref{fig:result_5ms_painlessmesh} ilustra os resultados de um dos experimentos. Na vertical são dispostos os sensores de S1 até S10. Na horizontal são apresentadas 3 barras referentes a troca de mensagens em 3 alturas distintas de voo dos drones, 20 metros, 35 e 50m. Todos os voos foram realizados a 5m/s.

Inicialmente é esperado que quanto mais perto dois rádios estivessem, mais dados eles poderiam trocar dada a intensidade do sinal. Ou, em contrapartida, um drone voando mais alto teria menos obstáculos entre ele e outro nó. Neste caso, quanto melhor fosse a linha de visada entre os rádios, mais dados seriam trocados entre os nós. 

O ponto interessante deste simples experimento foi notar que um dos nós nunca trocou dados com os demais e em algumas alturas nunca houve troca de dados. Analisando mais profundamente o que ocorria em dias e mais dias de testes no campo foi possível identificar que os nós não conseguiam fechar o enlace entre eles. Isso era difícil com os drones voando lentamente a 5m/s e nunca houve troca relevante com voos a 10m/s neste cenário específico. Ressaltasse que um dos nós nunca trocou dados, mesmo não estando em uma zona de sombra significante.

Todo o aparato de software para habilitar uma rede Mesh muitas vezes requeria um tempo superior ao \textit{handover} entre os nós. Se por um lado uma rede com infraestrutura seja inviável em diversos cenários, uma rede Mesh também pode não ser uma alternativa. De fato, há trabalhos que propõem protocolos MAC específicos para os \textit{rendevouz} de aeronaves\cite{Araghizadeh2016}\cite{Zafar2017}.

\subsection{Redes baseadas em Broadcast - 802.15.4}
\label{sec-sub:resultados-802.15.4}

Um outra abordagem para o experimento apresentado a cima na seção \ref{sec-sub:resultados-mesh} foi a troca do uso de 802.11s pelo uso de uma \textit{stack} 802.15.4. Neste caso a implementação utilizada foi a ESP-NOW\footnote{\url{https://docs.espressif.com/projects/esp-idf/en/latest/esp32/api-reference/network/esp_now.html}} que habilita um paradigma orientado a \textit{broadcast} com \textit{payloads} de até 250 bytes. 

O mesmo mapa, locais e drones foram utilizados e um ambiente idêntico também foi simulado. A Figura \ref{fig:result_5ms_802154} apresenta os resultados. Neste caso, todos os nós conseguiram trocar dados e em escala 10 vezes maior. Além das trocas de dados utilizando drones reais, são apresentados os resultados das simulações(SIM@) realizadas no GrADyS-SIM\cite{gradysSIMsbrc2022} em OMNET++/INET. Mais detalhes de como esses resultados foram utilizados para incrementar a simulação podem ser encontrados no trabalho de Olivieri et al. \cite{bolivieri2023}.

\subsection{802.11 Ad Hoc}
\label{sec-sub:resultados-wifi-adhoc}

Seguindo nossa linha de habilitar FANETs para a coordenação de enxames de drones, convergimos para a utilização de SBC que rodam Linux embarcados nos drones. Nesses SBCs (RaspBerry Pi 4) as trilhas ligavam os rádios ás antenas impressas nas placas foram removidas e conectores \textit{ipex} foram soldados\footnote{\url{https://youtu.be/MTwWnZG8wUY?si=MF3wNg9aAs3Mc5mi}}. Dessa maneira antenas externas ao SBC foram instaladas, antenas de roteadores WiFi de boa qualidade.

\begin{figure}
\caption{Dois drones montados por alunos com RaspBerry Pis}
  \label{fig-teste-dois-drones}
  \centering
  \includegraphics[width=1\textwidth]{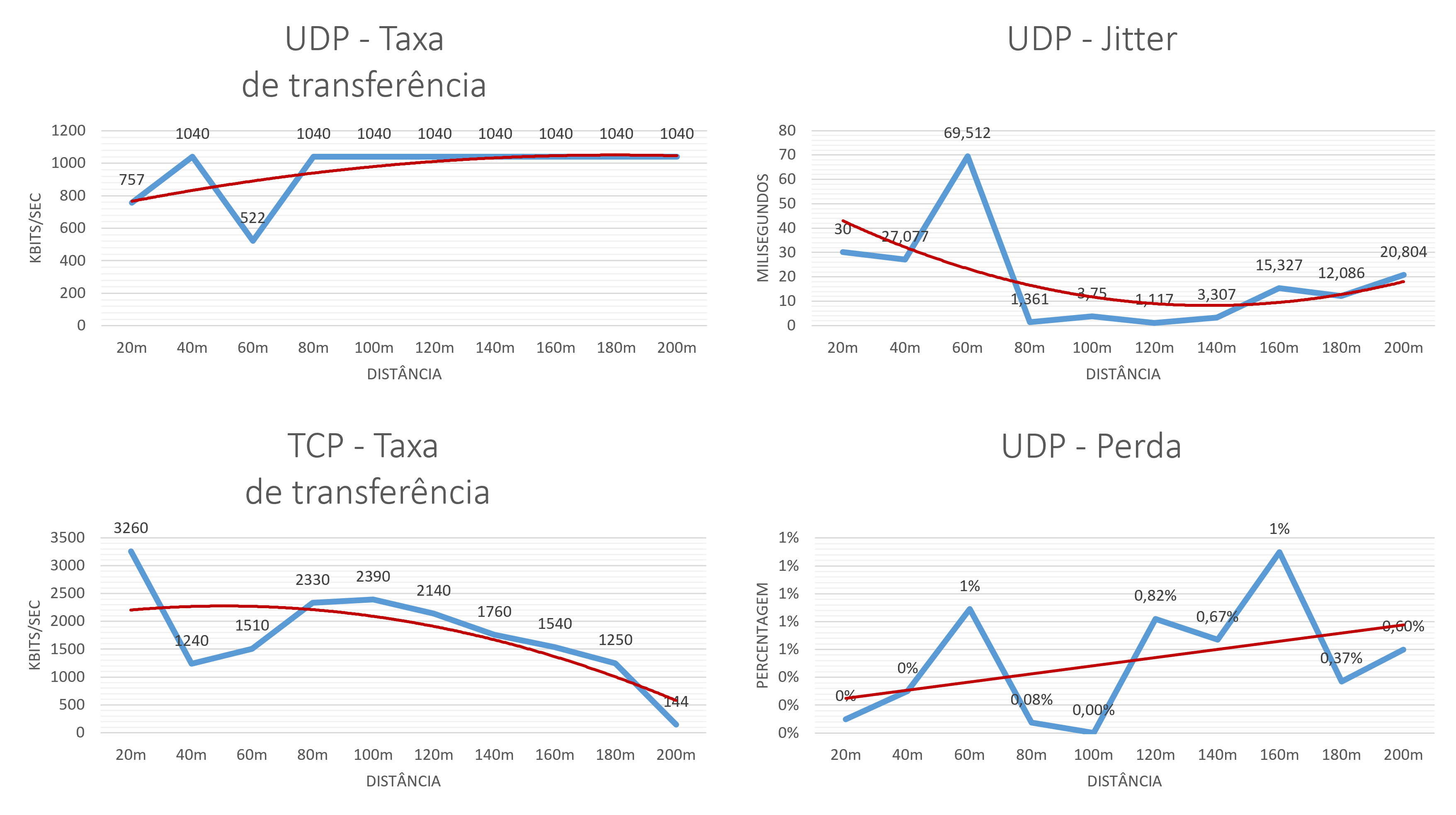}
\end{figure}

Uma rede 802.11 Ad Hoc foi configurada nos SBC que rodam Linux e podem ser seus device drivers ajustados quando necessário. Assim, uma rede com suporte a IP foi habilitada. O experimento aqui consistiu em testar o tráfego entre dois drones em distâncias incrementais. Um drone ficou parado a 10 metros de altura e outro foi se distanciando em intervalos de 20 metros até chegar a 200 metros. Em cada parada, o drone que se movia executava testes com os parâmetros \textit{default} da ferramenta iPerf3\footnote{\url{https://iperf.fr/}}. A Figura \ref{fig-teste-dois-drones} apresenta os resultados com uma linha vermelha representando uma curva de tendência de grau 2. Este experimento doi realizado 5 vezes e suas médias são apresentadas.

O primeiro resultado foi a observância das taxas de transferência utilizando TCP. Como esperado, conforme os drones se afastam as taxas de transferência diminuem. Caso a aplicação em teste não precise trocar algum dado volumoso como imagem ou som, isso pode não ser problema. Contudo, caso se necessite de maior robustez no link entre os drones, a utilização de UDP com um controle de fluxo pode ser mais resistente conforme a perda de pacotes aumenta. Regulando a taxa de transmissão UDP a 1Mbit/s, um link estável pode ser mantido em até 200 metros.

\section{Discussões e conclusões}
\label{sec:discussoes-clonclusao}

Simulações são fundamentais e agilizam muito o processo de pesquisa, seja de FANETS ou quaisquer propostas alinhadas com sistemas distribuídos. Contudo, testes de mundo real são imprescindíveis para Validar certas propostas. Ao se testar teses em campo, variáveis de ambiente alheias as simulações se fazem presente. Com essas novas variáveis a tese passa por um teste mais robusto e os simuladores podem ser incrivelmente melhorados como cita no artigo de referência na seção \ref{sec-sub:resultados-802.15.4}.

Para testes em mundo real, nenhum preparo pode ser subestimado. O número de pessoas envolvidas em um teste de campo sempre pode ser incrementado à fim de se ter maior controle dos experimentos. Os \textit{tradeoffs} das escolhas pelos equipamentos devem ser profundamente analisados e a ótica de reutilização e customização podem ser valorizadas. 

Este trabalho apresentou algumas visões e resultados de um projeto em franco andamento e espera gerar contribuições não só de pesquisa básica, mas também de pesquisa aplicada em FANETs em mundo real. Esperamos entregar a comunidade científica um conjunto de ferramentas reutilizáveis e experiências neste contexto.

Diversos trabalhos futuros podem ser analisados na melhoria das simulações de FANET com experimentos em campo. Neste projeto será priorizada a evolução \textit{pari passu} com o framework de simulação e com os softwares utilizados em campo a medida que pesquisas individuais prosseguem. 

\section*{Acknowledgments}
This study was financed in part by AFOSR grant FA9550-23-1-0136.

\bibliographystyle{IEEEtran}
\bibliography{2024-03-29-olivieri}

\end{document}